\documentclass[12pt,twocolumn]{iopart}
\usepackage{graphicx}
\usepackage{dcolumn}
\usepackage{bm}

\begin{document}

\title{Loading chromium atoms in a magnetic guide}

\author{A. Greiner, J. Sebastian, P. Rehme, A. Aghajani-Talesh, A. Griesmaier, T.Pfau}
 \ead{t.pfau@physik.uni-stuttgart.de}
 \address{5.~Physikalisches Institut, Universit\"{a}t Stuttgart.}
\date{\today}

\begin{abstract}
We have realized a magnetic guide for ultracold chromium atoms by
continuously loading atoms directly from a Zeeman slower into a
horizontal guide. We observe an atomic flux of $2\cdot10^7$atoms$/$s
and are able to control the mean velocity of the guided atoms
between $0\;$m$/$s and $3\;$m$/$s. We present our experimental
results on loading and controlling the mean velocity of the guided
atoms and discuss the experimental techniques that are used.
\end{abstract}
\submitto{\jpb}
\maketitle
\begin{figure}[htp]
\begin{center}
\includegraphics[width=8cm,keepaspectratio=true]{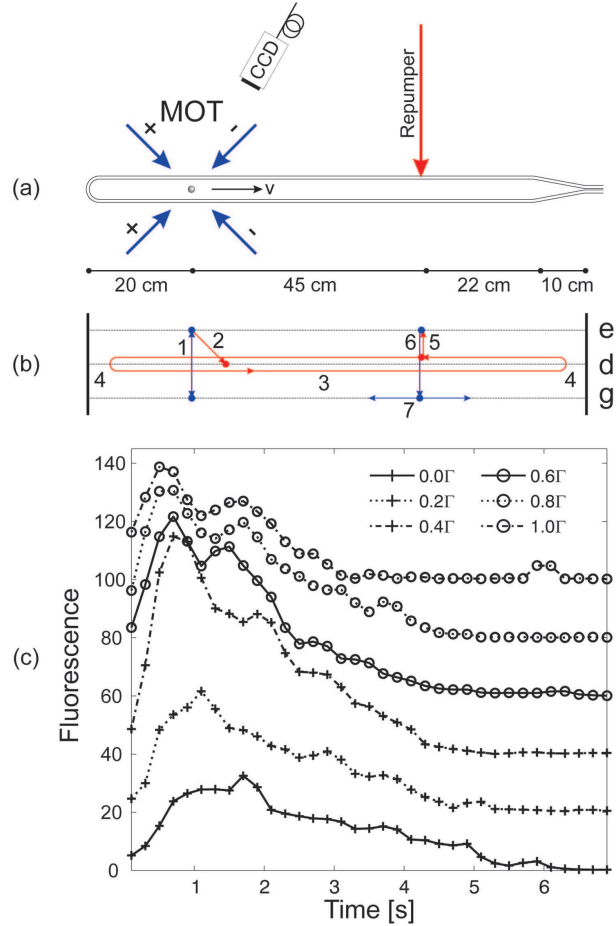}
\caption{\rm\fontsize{10}{12pt}\selectfont Setup for continuously detecting atoms in the guide (a) and chronology of the detection 
process (b). The data taken with the CCD-Kamera (c) show a strong dependence between the mean velocity 
$v=\frac{\epsilon}{2k\cos\theta}$ and the number of atoms.\label{fig:MOT_schema}}
\end{center}
\end{figure}

Experiments on ultracold atoms and in particular
Bose-Einstein condensation (BEC) of dilute gases~\cite{Streed:2005}
are currently limited by the pulsed operation in which they have to
be performed. However, since the achievement of Bose-Einstein
condensation and the first demonstration of a pulsed coherent beam
of matter waves (atom
laser)~\cite{Mewes:1997,Bloch:1999,Guerin:2006,Trippenbach:2000,Hagley:1999},
there is also the pursuit to generate a continuous wave
atom-laser~\cite{Hagley:2001,Chikkatur:2002}. Not only would such an
atom-laser have an immense impact on precision
measurements~\cite{Gustavson:1997,McGuirk:2002,Wildermuth:2005,Shin:2005},
but could also be a continuous source of atoms occupying the
smallest possible volume in phase space and thus allowing extreme
control over their external degrees of freedom. This could lead to
techniques that allow for much higher resolutions
and still considerable deposition rates in nanostructuring processes.\\
Current BEC experiments are carried out by performing several
cooling steps in a chronological sequence to increase the phase
space density. The reason is that magneto optical traps have to be
used to trap and cool the flux of atoms produced by thermal sources.
The phase-space densities, however, that are achievable in
magneto-optical traps, are limited to values orders of magnitude away
from degeneracy. The presence of the near resonant light used for
optical cooling prevents lower temperatures at the required spatial
densities, such that the light sources have to be switched off
before using alternative methods to increase the phase-space density
further. This also causes a disruption of the flux of atoms, leading
to the pulsed operation of these experiments. A source, however,
that would deliver a constant flux of atoms, already at temperatures
low enough to cool them with non-optical methods, could allow one to
perform such experiments in a continuous way. An important
prerequisite to achieve a continuous flux is to separate the cold
atoms spatially from the thermal sources and the laser cooling
region. The logical way to realize this is to cool the atoms
magneto-optically in a moving frame and to guide them away from the
thermal source and the magneto-optical trap after they have
interacted with the cooling light for some time. To reproduce the
subsequent cooling steps that are usually carried out in a temporal
sequence, the next cooling steps are then performed in a spatial
sequence as the atoms travel along the guide. To achieve a cw
atom-laser, degeneracy would have to be reached either within the
guide or in a three-dimensional trapping potential that is attached
to the end of the guide and is continuously loaded with ultracold
atoms from the guide~\cite{Roos:2003}.\\
Several groups have already demonstrated successful loading of
magnetic guides with ultracold atoms~\cite{Cren:2002,Teo:2002}. In
quasi-continuous operation, even the collisional regime could be
reached and thermalization of guided atoms was
observed~\cite{lahaye:2004}. We have realized a continuously loaded
magnetic guide by magneto-optical trapping of chromium atoms
directly from a Zeeman slower in the field of the magnetic guide.
Chromium atoms are especially well suited for such a continuous
loading scheme because of the large magnetic moment of $6\;\mu_B$ in
their $^7$S$_3$ ground state and a metastable state $^5$D$_4$ that
is decoupled from the cooling transition. Continuous loading of
chromium atoms has already been demonstrated for quadrupole and
cloverleaf traps~\cite{Stuhler:2001,Schmidt:2003a}. Due to their
large magnetic moment the necessary magnetic field gradients to
realize a guide are rather low, such that efficient operation of a
magneto-optical trap within the guide is possible. It has been shown
that the flux that is achievable with these loading schemes
is sufficient to reach quantum degeneracy~\cite{Griesmaier:2005a}.\\
Furthermore, chromium is a standard mask material in lithographic
processes and is also well suited for atom
lithography~\cite{Oberthaler:03a}. Direct deposition of
laser-focused thermal chromium atoms on a substrate has been used to
produce nanostructures~\cite{McClelland:93a,Drodofsky:97a} and
structured doping by simultaneously depositing a homogeneous matrix
material and laser-focused chromium~\cite{Schulze:01} has been
demonstrated. The generation of a continuous flux of ultracold atoms
in a continuously loaded magnetic guide is a first step on the way
from using incoherent thermal atomic beams to coherent matterwave
sources for atom litho\-graphy.\\
\begin{figure}
\begin{center}
\includegraphics[width=8cm,height=8cm,keepaspectratio=true]{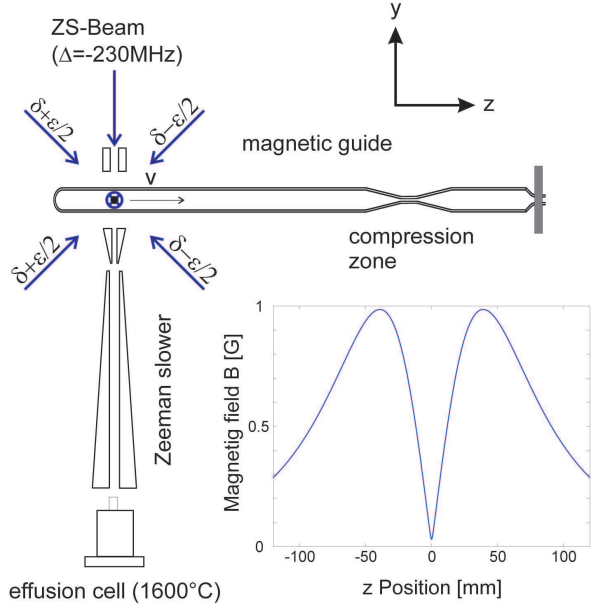}
\caption{\rm\fontsize{10}{12pt}\selectfont Schematic setup of our experiment. A MOT is loaded by atoms from an effusion cell 
slowed down by a Zeeman slower. The magnetic field gradient for the MOT is formed by 4 bars which also form the magnetic guide. By 
detuning the diagonal MOT beams, atoms are accelerated in the guide. An additional coil above the guide compensates the field of 
the last Zeeman slower coil. This creates an axial potential as shown in the plot.\label{fig:experiment_schema}}
\end{center}
\end{figure}
The setup that is used to load ultracold chromium atoms in our magnetic guide is shown schematically in 
Fig.~\ref{fig:experiment_schema}. A beam of chromium atoms is produced by a high temperature effusion cell operating at 
1600$^\circ$C which corresponds to a mean velocity of the atoms of 1000m/s. The pressure in the lower chamber that contains the 
effusion cell is about $10^{-7}$mbar. Since the capture velocity of our MOT is limited to 30m/s, a spinflip Zeeman slower is used 
to decrease the velocity of the atoms. Our 0.5m long Zeeman slower can decelerate atoms with velocities smaller than 400m/s. This 
Zeeman slower acts also as a differential pumping stage such that a pressure of below $10^{-10}$mbar can be maintained in the 
upper chamber where the trapping experiments are performed. The slow atoms are subsequently captured in the MOT. The magnetic 
guide is based on four water cooled Ioffe bars with a length of 1.5m and an distance of 4.6cm. A current of 170A provides a radial 
magnetic field gradient of 13G/cm. In the compression zone about 67cm away from the MOT region, the distance between the Ioffe 
bars is reduced to 9mm and the gradient is increased to 340G/cm. For atoms with a low velocity along the guide axis, the magnetic 
field of the compression zone acts as a magnetic mirror. The guide is closed on the other side by the round connection between the 
bars in the vacuum chamber, which also provides a reflecting field for slow atoms. In this sense cold atoms can be trapped between 
the two magnetic mirrors in the guide in a 80cm long trap. At a distance of about 45 cm from the MOT, a viewport can be used to 
observe the atoms in the guide (see Fig.~\ref{fig:MOT_schema}).\\
One advantage of chromium is the large magnetic moment in the ground state $^7S_3$ as well as in the metastable state $^5D_4$. In 
the extreme low field seeking states $|^7$S$_3,m_F=3>$ and $|^5$D$_4,m_F=4>$, the magnetic moment is $6\mu_B$. This makes sure 
that the atoms are held by the magnetic field of our MOT. The magnetic field needed to compensate gravitation in these states is 
about 1.7 G/cm. Therefore the atoms in the ground state  as well as the atoms in the metastable state can be loaded directly into 
our magnetic guide.\\
\begin{figure}
\begin{center}
\includegraphics[width=6cm,height=6cm,keepaspectratio=true]{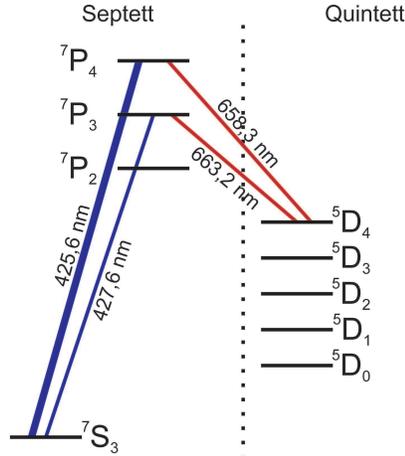}
\caption{\rm\fontsize{10}{12pt}\selectfont Part of the chromium level scheme. The cooling transition is $^7S_3$ to $^7P_4$. After 
about 250000 cycles on the cooling transition the chromium atoms decay to the metastable state $^5D_4$. The atoms can be pumped 
back to the $^7S_3$ ground state via the $^7P_3$ state. \label{fig:Termschema}}
\end{center}
\end{figure}
A part of the level scheme of chromium with the important levels for our experiment is shown in Fig.~\ref{fig:Termschema}. For 
cooling and slowing down the atoms we use the $^7S_3$ to $^7P_4$ transition at a wavelength of 426nm. This transition has a 
natural linewidth of $\Gamma=2\pi\cdot 5$MHz and a Doppler temperature of $124\mu K$. The transition $^7P_4$ to $^5D_4$ with a 
natural linewidth of $\Gamma=2\pi\cdot 20Hz$ leads to a continuous occupation of the $^5D_4$ metastable state inside the MOT after 
an average of 250000 cycles on the cooling transition. These atoms are not influenced by the light of the cooling transition and 
can move undisturbed inside the guide. To transfer the atoms back into the ground state, a laser that pumps the atoms from the 
$^5D_4$ to the $^7P_3$ state at a wavelength of 663.2nm is used. From the $^7P_3$, the atoms decay with high probability to the 
ground state.\\
We use a resonant probe beam overlapped with a repumping beam to detect atoms in the guide. Metastable atoms that cross the 
repumping beam are transferred to the ground state via the excited state. Due to the presence of the resonant probe beam, atoms in 
the ground state can emit multiple photons by resonant scattering before they are expelled from the detection region. We detect 
emitted fluorescence photons with a photomultiplier tube in front of which we have placed an optical filter to block the 663 nm 
light of the repumper. Modulating the intensity of the repumper by an AOM allows us to use a lock-in detection method to suppress 
background noise.\\
The 425nm laser light used for the MOT and Zeeman slower is produced by intracavity second harmonic generation of infrared light 
from a Ti:Sapphire laser. The laser is pumped by a frequency doubled diode pumped solid state laser system and delivers a total of 
2.8W laser power at 851nm. An LBO crystal is used as the non-linear medium and the home-made bow-tied cavity is locked using the 
H\"{a}nsch-Couillaud-scheme \cite{Hansch:1980}. The total power of 426nm light produced by the cavity amounts up to 1.3W. The 
frequency stabilization of the cooling transition is assured by locking the laser to the errorsignal obtained by performing 
Doppler free polarization spectroscopy chromium in a hollow cathode lamp. About 4mW of repumper light at a wavelength of 663 nm is 
produced by a home made laserdiode system. The frequency of this system is stabilized on a resonator which is locked to the 
frequency of the Ti:Sapphire laser.\\
The MOT is formed by 5 beams (Fig.~\ref{fig:experiment_schema}). Four of them are diagonal to the axis of the magnetic guide. 
Their frequencies ca be controlled individually and they are used to cool and accelerate the atoms in the guide. The fifth beam is 
orthogonal to the axis of the magnetic guide and retroreflected.  For an independent control of the frequency of the MOT beams, 
acousto optical modulators (AOM) in double pass configuration are used. To provide a Gaussian mode for the MOT beams and to 
prevent that vibrations disturb the stability of the MOT, the light is brought to the vacuum chamber via glass fibers and the 
output couplers are directly mounted at the viewports. The typical power of the MOT beams is 20mW per beam. By changing the 
frequencies of the diagonal MOT beams with respect to each other, one can cool the atoms in a moving frame. The velocity of the 
atoms is given by \cite{Cren:2002}
\begin{eqnarray}
v=\frac{\epsilon}{2k\cos\Theta}\; ,
\label{eq:mm_mot}
\end{eqnarray}
where $\epsilon$ is the detuning between the beams (see Fig.~\ref{fig:MOT_schema}), $\Theta$ is the angle between the propagation 
direction and the direction of the MOT beams, and $k$ is the wavevector of the light. For our cooling transition at 426nm a 
detuning of $\epsilon=1\Gamma$ corresponds to a velocity of about $1.5\frac{m}{s}$.\\
For a high loading rate in the MOT, the end of the Zeeman slower should be as close to the MOT as possible. A drawback, however, 
is that the last Zeeman slower coil leads to a residual magnetic field in the MOT region that shifts the zero of the magnetic 
guide in the y direction. To compensate this shift, a coil that produces a compensation field is used opposite to the last Zeeman 
slower coil. These two coils form a 3D magnetic trap with a gradient of about 0.25G/cm. Since the Ioffe bars produce a gradient of 
13G/cm, the influence of this field in the x and y direction is negligible. However, along the  center of the guide, these coils 
produce a potential as shown in the inset in Fig.~\ref{fig:experiment_schema}. The maximum of this field is about 1Gauss which 
corresponds to a trap depth of $400 \mu K$ and is a barrier for ultracold atoms.\\
To characterize the parameters of our system, we first measured the loading rate and lifetime of the chromium atoms in our 
magnetic guide. To obtain the numbers of atoms in the guide, we used the repumper at the transition $^5D_4$ to the $^7P_3$ state 
from which they decay in the ground state by emitting exactly one photon of 427.6nm. By counting these photons with the help of a 
photomultiplier, one can deduce the calibrated atomflux. We measured a continuous loading rate of $2\cdot 10^{7}$ atoms per second 
with this method. The lifetime of the chromium atoms in the magnetic guide was measured to be 68.3s. The influence of the hot 
atoms from the oven in the MOT zone was also measured and leads to a decrease of the lifetime to 44.6s.\\
To analyse the dependence of the mean velocity of the atoms on the detuning of the MOT qualitatively, we used the scheme that is 
shown schematically in Fig.~\ref{fig:MOT_schema}~(b) (numbers in parentheses correspond to the respective numbers in 
Fig.~\ref{fig:MOT_schema}~(b)). The atoms are continuously cooled by the MOT-beams (1) and leave the MOT in the metastable state 
(2). The guide is filled up with these atoms (3) while the compression zone and the curved part of the Ioffe-Bars (4) act as 
reflecting magnetic mirrors. After loading the guide for 120s, all the MOT beams except for one diagonal MOT beam are put off and 
the repumper in the MOT zone is put on for 100ms. Thereby all atoms inside the magnetic trap produced by the 4 Ioffe Bars and the 
residual field of the Zeeman slower and the compensation coil are removed. After this removing step, the repumper in the detection 
zone is put on for 500ms (5). Thus atoms that pass the detection zone are pumped back to the ground state (6) and leave this zone 
in their current direction of propagation. Depending on this direction and their absolute velocity, the atoms need different times 
to travel along the guide before they arrive in the MOT region where they are detected by the help of the diagonal MOT beams. For 
narrow velocity distributions ($\frac{\Delta v}{v}$) one expects two peaks arriving at the MOT region, while for broad velocity 
distributions the two peaks should smear out. Fig.~\ref{fig:MOT_schema}~(c) shows the obtained data that show the expected 
behaviour. In addition, one can see an increase of the number of atoms for higher velocities. This can be explained by the fact, 
that the atoms need a certain velocity to climb up the magnetic potential formed by the last Zeeman slower coil and the 
compensation coil. Although a quantitative analysis of the velocity distribution was not made due to the complication in the MOT 
region, these data show a velocity dependence of the atoms for different detunings and hence show the first measurement of 
continuously loaded chromium atoms into a magnetic guide.\\
\begin{figure}
\begin{center}
\includegraphics[width=8cm,keepaspectratio=true]{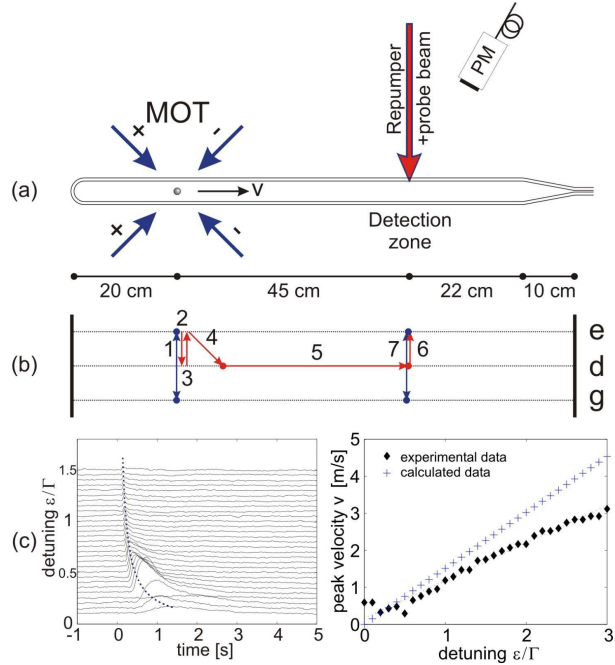}
\caption{\rm\fontsize{10}{12pt}\selectfont Setup for a velocity dependent detection of the atoms (a) and chronology for the used 
sequence (b). The data were obtained by measuring the fluorescence light with a photomultiplier. The dependency between velocity 
and detuning was calculated (c).  \label{fig:Guide_schema}}
\end{center}
\end{figure}
Figure \ref{fig:Guide_schema} presents a quantitative study of the dependence between the detuning of the MOT and the velocity of 
the atoms transferred into the guide. The data was taken by launching packets of atoms at t = 0 from the MOT into the guide and by 
observing a fluorescence signal produced by the atoms at some distance from the MOT as sketched in 
Fig.~\ref{fig:Guide_schema}~(a). The loading and detection sequence can be represented by the state diagram shown in 
Fig.~\ref{fig:Guide_schema}~(b) (numbers in parentheses correspond to the respective numbers in Fig.~\ref{fig:Guide_schema}~(b)). 
(1+2) We first load the MOT for 5 s without any relative detuning. This leads to an accumulation of metastable atoms in the MOT 
region, where the residual magnetic fields of the Zeeman slower and its compensation coil together with the Ioffe bars produce a 
three dimensional magnetic trap. This accumulation significantly increases the total number of atoms present in a single shot and 
thus leads to an enhancement of the signal that we want to measure. Since this trap is rather shallow, a certain fraction of the 
metastable atoms can leak from the MOT region into the magnetic guide during the loading of the MOT. These atoms would create 
undesired background fluorescence during our measurement. We therefore shut off the MOT beams after loading the MOT and wait until 
all atoms in the guide are eliminated by the light force of the beams in the detection region. In order to transfer atoms with a 
certain velocity into the guide, we turn on the MOT beams at t=0 with a preset detuning $\epsilon$. Simultaneously, we switch on a 
repumping beam that crosses the center of the MOT.
(3) The repumper transfers the trapped metastable atoms into the ground state where they are subject to the detuned MOT beams. (4) 
As a result, these atoms acquire a velocity proportional to the detuning of the MOT beams. (5) Atoms in the MOT cycle that decay 
back into the metastable state are guided by the magnetic potential of the Ioffe bars and are thus able to escape from the MOT 
region into the magnetic guide.
(6+7) The fluorescence light of the atoms generated by a resonant probe beam overlapped with a repumping beam is detected with the 
help of a photomultiplier about 45cm away form the MOT zone.
In the left plot of Fig.~\ref{fig:Guide_schema}~(c), the fluorescence signal observed for different detuning is plotted over time. 
One can see that with increased detuning, the guided atoms arrive faster at the detection region. This is depicted in more detail 
in the right plot of Fig.~\ref{fig:Guide_schema}~(b) where the peak velocity of each of the velocity distributions of the left 
plot is plotted as a function of $\epsilon$. The data show the expected linear dependence of the peak velocity on the MOT 
detuning. The value obtained for the constant of proportionality differs slightly, however, from the expected theoretical 
behaviour represented by the blue dotted line. Whether this deviation is due to the complex field and polarisation configuration 
in the MOT region is subject of further investigations.\\
We have demonstrated the successful continuous loading of ultracold chromium atoms in a magnetic guide. A flux of 
$2\cdot10^7$atoms/s in the guide was achieved. We were able to vary the mean velocity of the guided atoms by detuning the beams of 
our MOT with respect to each other. We have studied the dependence of the mean velocity on this detuning. We find the expected 
linear dependence between the detuning and the velocity. Mean velocities  of up to $3\;$m/s have been achieved. These velocities 
should be sufficient for the atoms to enter the tapered region of the guide where we plan to perform further Doppler cooling in 
order to increase the phase space density of the guided atoms.\\
The work presented in this paper was supported by the Landesstiftung Baden-W\"{u}rttemberg and the German-Israeli Foundation. 
A.A.-T.~acknowledges financial support by the Studienstiftung des deutschen Volkes.

\end{document}